\documentclass[conference]{IEEEtran}
\IEEEoverridecommandlockouts
\usepackage{cite}
\usepackage{amsmath,amssymb,amsfonts}
\usepackage{algorithm}
\usepackage{algpseudocode}
\usepackage{graphicx}
\graphicspath{{resultfigs/}{modelfigs/}}
\usepackage{textcomp}
\usepackage{xcolor}
\usepackage{hyperref}
\usepackage{booktabs}
\usepackage{enumerate}
\usepackage{enumitem} %

\usepackage{siunitx} 

\usepackage{siunitx}
\usepackage{tabularx}

\usepackage{pifont}

\usepackage{multirow}

\usepackage{hyperref}

\usepackage{tikz}

\usepackage{makecell}

\def\BibTeX{{\rm B\kern-.05em{\sc i\kern-.025em b}\kern-.08em
    T\kern-.1667em\lower.7ex\hbox{E}\kern-.125emX}}
\begin{document}

\title{Covertness Recovery in Multi-RAT UAV Swarms Under Post-Detection Threat Escalation via Lifelong Reinforcement Learning
\thanks{Research was sponsored partially by the US National Science Foundation under Grant CNS-2008309 and the Army Research Laboratory and was
accomplished under Cooperative Agreement Number W911NF-23-2-0014.
The views and conclusions contained in this document are those of the
authors and should not be interpreted as representing the official policies,
either expressed or implied, of the Army Research Laboratory or the U.S.
Government. The U.S. Government is authorized to reproduce and distribute
reprints for Government purposes notwithstanding any copyright notation
herein. The authors also thank NVIDIA Corporation for providing equipment used in this research.}
}

\author{
\IEEEauthorblockN{Yinxuan Wu, Xiaohao Xia, Beatriz Lorenzo, Dennis Goeckel}
\IEEEauthorblockA{\textit{ECE}, \textit{University of Massachusetts Amherst}, Amherst, MA, USA \\
\{yinxuanwu,xiaohaoxia,blorenzo,dgoeckel\}@umass.edu}
\and
\IEEEauthorblockN{Fikadu Dagefu}
\IEEEauthorblockA{\textit{US Army Research Laboratory}, Adelphi, MD, USA \\
\{fikadu.t.dagefu.civ\}@army.mil}}



\maketitle


\begin{abstract}
Unmanned aerial vehicle (UAV) swarms increasingly span multiple spectrum bands and radio access technologies (RATs), e.g., Wi-Fi, LoRa, Zigbee, and cellular. With active adversaries capable of protocol-aware jamming, spoofing, traffic analysis, and control-plane exploitation, survivable and timely covert communication is challenging. 
\textcolor{black}{Moreover, after detection, adversaries may escalate with selective, adaptive attacks that increase latency and hinder recovery. }
This paper is the first to study covertness recovery under post-detection threat escalation. We design a learning-based mechanism that re-establishes covert communication across learning domains with different Willie densities and attack models, amid stochastic task arrivals, UAV mobility, and uncertain warden locations and behaviors. We jointly optimize RAT selection, resource allocation, and task offloading to minimize Age of Covert Information (AoCI), energy, and threat impact. Warden behavior is modeled via an adaptive attack-selection process capturing detection-triggered escalation and nonstationary strategies. We propose Cross-Domain Factorized Adaptive Lifelong Learning with Proximal Policy Optimization (CDFALL-PPO), which factorizes shared and domain-specific policy components for online adaptation while mitigating catastrophic forgetting.
\textcolor{black}{Across three escalating 30-task sequences, extensive simulations show that CDFALL-PPO raises median packet covert delivery rate (PCR) by 14.5 \%, increases covert throughput by up
to 53.6\%, and lowers AoCI by 59.4\% relative to PPO.  
Under the strongest composite attack,} 
mean on-attack PCR improves from 52\% to 70\%, and \textcolor{black}{mean post-attack recovery PCR rises from 64\% to 93\%.}

\end{abstract}


\section{Introduction}
UAV-assisted networks increasingly operate over multiple spectrum bands and heterogeneous RATs, 
to support mission-critical sensing, control, and data offloading.
Such applications impose stringent privacy and security requirements, where transmissions should remain undetectable to wardens attempting to infer communication.
This paradigm is known as covert communication, whose information-theoretic foundations were initiated by Bash et al.~\cite{6584948} and have since been extensively studied and extended in the literature~\cite{10090449}.

For UAV-assisted covert communications, 
Xia et al.~\cite{10757651} studied UAV covert communications in heterogeneous Internet of Things (IoT) and optimized cross-technology data offloading under passive wardens.
Wu et al.~\cite{11363010} investigated the joint UAV positioning, transmit power, and blocklength optimization to minimize secure age of information. They solve the problem by using a novel convex approximation.
Qian et al.~\cite{11368746} modeled a Bayesian Nash game between the transmitter and warden and use stochastic geometry to derive interference and outage for a UAV-relayed Rayleigh channel. They prove equilibrium existence and show it achieves the optimal covert rate.
Yang et al. \cite{10418490} considered UAV covert communications against colluding active wardens, which can actively probe or interfere with transmissions rather than only eavesdrop, under location uncertainty. They jointly optimized the UAV’s 3D trajectory and transmit power to maximize covert rate under worst-case covertness constraints. However, these works primarily optimize covertness and throughput (or AoI) under a fixed threat model, and do not explicitly address how to recover covert communications after detection when the warden escalates its interference strategy.

UAV swarms are exposed to diverse security threats, motivating explicit threat modeling for UAV-assisted wireless networks~\cite{10.1145/3703625}. Prior work spans physical-layer spoofing detection using channel state information (CSI)-based feature learning and subspace methods for pilot spoofing identification~\cite{10680118,10666028}, as well as impersonation attacks and defenses that mimic legitimate devices or undermine angle-of-arrival (AoA)-based authentication~\cite{10694672}. Denial-of-service (DoS) and distributed DoS (DDoS) threats have also been addressed via learning-based detection and mitigation in UAV-enabled architectures~\cite{10634504}.
These challenges are particularly pronounced in covert multi-RAT systems, 
where a protocol-aware warden can escalate from passive detection
to coordinated disruption once covert activity is detected~\cite{11026809}. 
This motivates further research on resilient covert communications that explicitly model post-detection threat escalation and enable rapid recovery while maintaining covertness across heterogeneous RATs.


\textcolor{black}{
Continual and cross-domain RL can accelerate adaptation in non-stationary networks~\cite{10906634}. Such long-term self-adaptation requires systems to continuously acquire, update, and reuse knowledge over time, which is a capability central to continual or lifelong learning (CL/LL).
Existing CL/LL studies include a deep lifelong learning framework with dictionary learning (DC-ELLA) for UAV-assisted Metaverse services~\cite{11396413} and semantic-aware UAV-swarm learning architectures~\cite{10807275}, early cross-domain LL via shared knowledge repositories~\cite{ammar2015autonomous}, and surveys on cross-domain transfer in reinforcement learning (RL)~\cite{10614179}.
However, they do not address CL/LL under active, strategic adversaries or explicitly mitigate forgetting under covert communication constraints. This is substantially more challenging as the environment evolves strategically in response to the agent’s actions, and learning must jointly optimize communication utility and detectability while reusing knowledge 
without exposing patterns that increase vulnerability.
}

Motivated by the above observations, this paper makes the following contributions: 
\begin{enumerate}[leftmargin=1.0em]
\item We formulate covertness recovery in a multi-RAT UAV swarm 
under Willie post-detection escalation, 
capturing adaptive attack selection over DoS, impersonation of Bob, bandwidth hijacking, 
and ACK-related control-plane knowledge acquisition, 
while the swarm controller executes recovery strategies to restore covert communication. 
\item We jointly optimize offloading, bandwidth allocation, transmit power, and interface re-initialization under covert constraints. 
The resulting problem is cast as a partially observable Markov decision process (POMDP), and we develop CDFALL-PPO, which factorizes the actor head to 
\textcolor{black}{enable lifelong adaptation across domains characterized by different map sizes, Willie coverage, and attack portfolios.}
\item Extensive experiments on three escalating 30-task sequences, each averaged over three random seeds to account for training stochasticity, show that CDFALL-PPO consistently 
achieves substantially stronger post-attack recovery even under composite Willie attacks.
\end{enumerate}

\begin{figure}
    \centering
    \includegraphics[width=0.95\linewidth]{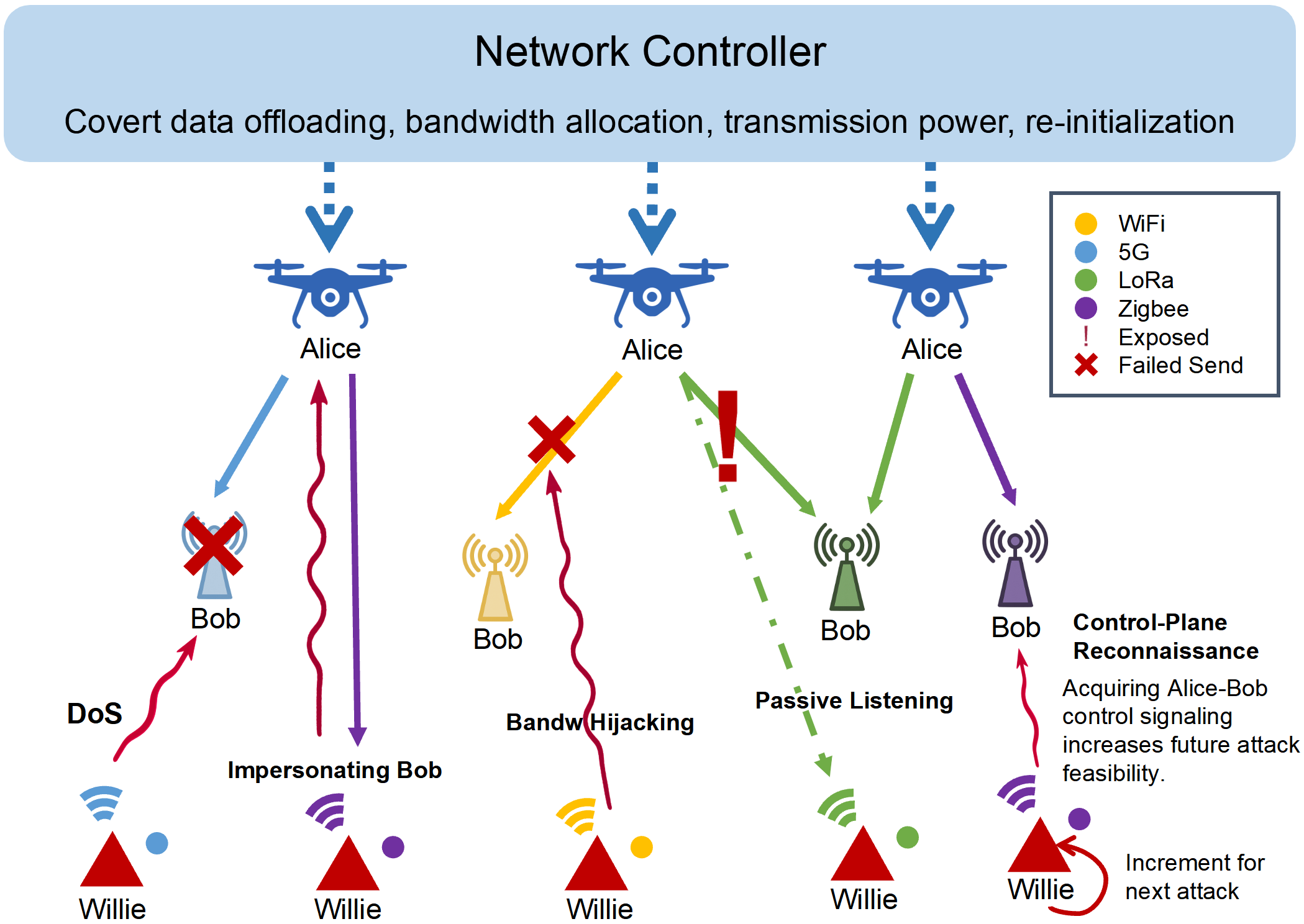}
    \caption{Multi-RAT UAV swarm under multiple threats 
    (DoS, impersonation of Bob, bandwidth hijacking, passive detection, 
    control-plane knowledge acquisition), 
    and knowledge increment for increasing future attack feasibility.}
  \label{fig:system_model}
\end{figure}


\section{System Model}
\label{sec:systemmodel}

We consider a heterogeneous UAV swarm of $M$ UAVs (Alices) $\mathcal M=\{1,\dots,M\}$ that collect IoT data and offload it to $J$ access points (Bobs) $\mathcal J=\{1,\dots,J\}$. Each Alice follows a 3-D random mobility model \cite{8671460}, is equipped with multiple radio interfaces, and can operate over $Z$ protocols $\mathcal Z=\{1,\dots,Z\}$. In contrast, each Bob $j$ is assumed to operate on a single protocol $z(j)\in\mathcal Z$, which models protocol-specific access infrastructure. IoT packet arrivals are modeled as a Poisson process with average rate $\lambda_i^{z}$. A set of $X$ ground wardens (Willies) $\mathcal X=\{1,\dots,X\}$ monitor Alice--Bob transmissions, where each Willie $x$ listens on a single protocol $z(x)\in\mathcal Z$, reflecting practical sensing constraints in which a warden monitors one RAT at a time. Let $\mathcal J_z$ and $\mathcal X_z$ denote the subsets of Bobs and Willies associated with protocol $z$, respectively. We assume that each Alice can obtain transmission feedback indicating whether a transmission remained covert through established indirect, side-channel, or ACK-based mechanisms~\cite{bash2013limits,goeckel2016lpd,10695865,yang2025survey,pi2025uavthz}. Modeling the feedback inference process is beyond the scope of this paper.

\subsection{Communication Model}
\label{sec:comm_model}
 We adopt a slotted system where each Alice $i \in \mathcal{M}$ collects data on interface $z \in \mathcal{Z}$ and offloads it to a selected Bob $j \in \mathcal{J}$ either on the same interface or, if cross-technology offloading is enabled, on another interface $z'$.  
For a transmission from Alice $i$ to Bob $j$ on protocol $z$, the achievable rate is
\begin{equation}
R_{i, j}^z(t)
=
b_{i,j}^z(t) B^z\log_2\bigl(1+ \rho_{i, j}^z(t)\bigr), \,\,
\rho_{i, j}^z(t)
=
\frac{P_{\mathrm{rx},i\to j}^z(t)}{N_0^z + I_j^z(t)}.
\label{eq:rate}
\end{equation}
where $b_{i,j}^z(t) \in[0,1]$ is the normalized bandwidth allocated of the total bandwidth $B^z$ on protocol $z$.
The received power is $P_{\mathrm{rx}, i\to j}^z(t) = P_{\mathrm{tx},i}^z(t) G_{\mathrm{tx},i}^z G_{\mathrm{rx},j}^z 10^{-\mathrm{PL}_{i, j}^z(t)/10}$, where $P_{\mathrm{tx},i}^z(t)$ is the transmit power, and $G_{\mathrm{tx},i}^z$, $G_{\mathrm{rx},j}^z$ are the linear antenna gains at the transmitter and receiver. 
$\mathrm{PL}_{i, j}^z(t)$ and $N_0^z$ are the path loss \cite{3GPP_TR_38901} and the thermal noise, and the interference from other Alice $i'$ at receiver $j$ is
$I_{j}^{z}(t)=\!\!\sum_{i'\neq i} P_{\mathrm{rx},i'\!\to\!j}^z(t)$.

Each Willie $x \in \mathcal{X}_z$ monitors protocol $z$ and performs a binary hypothesis test between $\mathcal H_0$ (Alice inactive) and $\mathcal H_1$ (Alice active). Using an energy detector over $N$ i.i.d.\ complex baseband samples per slot, and conditioned on the instantaneous noise-plus-interference power $N_x^z(t)$ and, under $\mathcal H_1$, Alice's received power $P_{\mathrm{rx},i\to x}^z(t)$, the observation vector is modeled as $\mathbf{y}_x^z(t)\!\sim\!\mathcal{CN}(\mathbf{0},\sigma^2_k \mathbf{I}_N)$ for $k\in\{0,1\}$, where $\sigma^2_0 = N_x^z(t)$ and $\sigma^2_1 = N_x^z(t)+P_{\mathrm{rx},i\!\to\!x}^z(t)$ with $N_x^z(t)$ denoting Willie's instantaneous noise-plus-interference power. To guarantee $\epsilon$-covertness for $\epsilon\in(0,1)$, we bound Willie's total detection error probability via Pinsker's inequality, which yields the KL-divergence constraint $D_{\mathrm{KL}}^{(N)}(\mathcal H_0\|\mathcal H_1) < 2\epsilon^{2}$. For the above $N$-sample complex Gaussian distributions, this reduces to $N \cdot \Bigl[\ln(1 + \rho_x(t)) - \frac{\rho_x(t)}{1 + \rho_x(t)}\Bigr] < 2\epsilon^{2}$, where $\rho_x(t) \triangleq P_{\mathrm{rx},i\!\to\!x}^z(t) / N_x^z(t)$ is Willie's instantaneous SINR. Since the left-hand side is strictly increasing in $\rho_x(t)$ for $\rho_x(t)\ge 0$, there exists a unique threshold $\rho_\epsilon$ such that covertness holds if and only if $\rho_x(t) < \rho_\epsilon$. Equivalently, to ensure covertness, the transmit power must satisfy
\begin{equation}
    P_{\mathrm{rx},i\to x}^z(t) < \rho_\epsilon N_x^z(t).
    \label{eq:cov_power}
\end{equation}

\subsection{Willie Attack Types, Feasibility, and Selection Policy}
\label{subsec:willie_attacks}

Each Willie $x\in\mathcal{X}_z$ continuously performs passive detection on its monitored protocol $z$. In each slot, Willie may attempt at most one active attack from $\mathcal K=\{ \mathrm{DoS\_B},$ $\mathrm{PH\_IB},$ $\mathrm{BH\_IA},$ $\mathrm{CPB\_ACK}\}$, where
$\mathrm{CPB\_ACK}$ is a reconnaissance action that improves Willie's ability to detect and decode Bob$\!\to$Alice ACK-related control signaling (e.g., ACK timing or resource allocation), thereby facilitating subsequent impersonation attacks~\cite{ieee802112020,ieee8021542020,lorawanSpec}. 
$\mathrm{DoS\_B}$ is a Bob-targeted DoS attack causing packet losses and retransmissions. 
$\mathrm{PH\_IB}$ is a packet hijacking attack where Bob impersonation at the link layer causes Alice to transmit packets intended for Bob to Willie instead. 
$\mathrm{BH\_IA}$ is a link-layer Alice-impersonation attack 
that hijacks bandwidth and increases AoCI.
\textcolor{black}{We assume the impersonation attacks $\{ \mathrm{PH\_IB}, $$\mathrm{BH\_IA} \}$ target a specific Alice interface $z$ and require Willie to recognize that interface.
The Bob-side attacks $\{ \mathrm{CPB\_ACK}, $$\mathrm{DoS\_B} \}$ may be attempted at any slot. }
Conversely, Alice may re-initialize interface $z$, denoted by $r_i^z(t)\in\{0,1\}$. Interface re-initialization refreshes the protocol context and can invalidate Willie’s interface-specific knowledge.  
For compactness, the following dynamics are described for a fixed candidate Alice $i$ and interface $z$; hence, we suppress their indices without ambiguity.

\textit{a) Willie attack feasibility:}
Let $\chi(t)\in\{0,1\}$ denote whether the current target interface is recognized by Willie, and let $\eta(t)\in[0,1]$ denote the ACK/control knowledge accumulated for that interface. Their dynamics are
\[
\small
\chi(t+1)=
\begin{cases}
0, & r(t)=1,\\
1, & \text{Willie decodes a packet in slot } t,\\
\chi(t), & \text{otherwise},
\end{cases}
\]
and 
\[
\small
\eta(t+1)=
\begin{cases}
0, & r(t)=1,\\
\min\{1,\eta(t)+\delta\}, & \mathcal E^{\mathrm{CPB}}(t),\\
\eta(t), & \text{otherwise}.
\end{cases}
\]
where  $\delta$ is Willie's knowledge increment, and 
$\mathcal E^{\mathrm{CPB}}(t)\triangleq\{A(t)=\mathrm{CPB\_ACK},\,I_{\mathrm{CPB}}^{\mathrm{imp}}(t)=1\}$ denotes the event that Willie launches a CPB\_ACK attack at slot $t$ and the impersonation attempt $I_{\mathrm{CPB}}^{\mathrm{imp}}(t)$ is successful.

\textcolor{black}{
Let $\bar p_{a,z}^{\mathrm{fea}}\in[0,1]$ denote the attempt
base feasibility probability of attack $a$ on protocol $z$ (conditioned on selecting attack $a$). 
}
The attempt
feasibility of attack $a$ at slot $t$ is
\begin{equation}
\small
p_{a}^{\mathrm{fea}}(t)=
\begin{cases}
\bar p_{a,z}^{\mathrm{fea}}, &
a\in\{\mathrm{CPB\_ACK},\mathrm{DoS\_B}\},  \\
\chi(t)\min\!\left\{1,\bar p_{a,z}^{\mathrm{fea}}+\eta(t)\right\}, &
a\in\{\mathrm{PH\_IB},\mathrm{BH\_IA}\}.
\end{cases}
\label{eq:p_fea_interface_final}
\end{equation}

Conditioned on selecting attack $a$, the implementation indicator is
\(
I_{a}^{\mathrm{imp}}(t)\sim
\mathrm{Bernoulli}\!\left(p_{a}^{\mathrm{fea}}(t)\right).
\)
For a window length $W$, the empirical conditional attack hit rate is
\textcolor{black}{
\begin{equation}
p_{a}^{\mathrm{hit}}(t)
\triangleq
\frac{\sum_{\tau=t-W+1}^{t}\mathbf{1}\{A(\tau)=a,\,I_{a}^{\mathrm{imp}}(\tau)=1,\,S(\tau)=1\}}
{\sum_{\tau=t-W+1}^{t}\mathbf{1}\{A(\tau)=a,\,I_{a}^{\mathrm{imp}}(\tau)=1\}},
\label{eq:pk_hit}
\end{equation}
}
where $A(\tau)$ is the selected attack and $S(\tau)\in\{0,1\}$ indicates 
\textcolor{black}{whether the implemented attack achieves its intended effect.} 
We set $p_a^{\mathrm{hit}}(t)=0$ when the denominator is zero.

We define the effective conditional success probability as
\begin{equation}
p_{a}^{\mathrm{eff}}(t)
\triangleq
p_{a}^{\mathrm{fea}}(t)\,p_{a}^{\mathrm{hit}}(t),
\label{eq:pk_eff}
\end{equation}
\textcolor{black}{
which is the probability that attack $a$ is both implementable and effective. If $\pi_a(t)$ is Willie's selection probability, then the \emph{slot-level marginal} probability that attack $a$ is selected and succeeds is $\Pr[A(t)=a,S(t)=1]=\pi_a(t) p_a^{\mathrm{eff}}(t)$.
}

\textit{b) Probabilistic attack-selection policy:}
\textcolor{black}{Willie trades off exploiting current capabilities and acquiring new knowledge. Let $\mathcal K_{\mathrm{exp}}\triangleq\{\mathrm{DoS\_B},\mathrm{PH\_IB},\mathrm{BH\_IA}\}$ denote the exploitation actions, and let CPB\_ACK be the knowledge-acquisition action. }
We define the current effective utility which measures the current effective exploitation capability  $S_{\mathrm{eff}}(t) \triangleq \sum_{k\in\mathcal K_{\mathrm{exp}}} p_{k}^{\mathrm{eff}}(t)$ and the unrealized potential due to limited feasibility (i.e., benefit of acquiring more knowledge), $G(t) \triangleq \sum_{k\in\mathcal K_{\mathrm{exp}}} p_k^{\mathrm{hit}}(t)\,[1 - p_k^{\mathrm{fea}}(t)]$.
The probability of choosing CPB\_ACK and executing an exploitation attack $k \in \mathcal K_{\mathrm{exp}}$ is:
\begin{equation}
\pi_{\mathrm{CPB\_ACK}}(t) = \frac{G(t)}{G(t)+S_{\mathrm{eff}}(t)}, \, \pi_k(t) = \frac{p_k^{\mathrm{eff}}(t)}{G(t)+S_{\mathrm{eff}}(t)}.
\label{eq:Willie_attack_select}
\end{equation}
If $G(t)+S_{\mathrm{eff}}(t)=0$, Willie selects an active attack uniformly at random.

\section{\textcolor{black}{Covert Data Offloading Optimization Framework}}
\label{sec:datasched}

We assume that a network controller (NC) manages the UAV swarm and jointly controls interface re-initialization, cross-technology offloading, Alice-Bob association, bandwidth, and power to maximize covertness while limiting AoCI, exposure, and energy cost.

\subsection{Decision Variables}
Let $y_{ij}^{z}(t)\in\{0,1\}$ indicate whether Alice $i$ transmits to Bob $j$ on interface $z$ in slot $t$. Each interface can be associated with at most one Bob per slot,
\begin{equation}
\sum_{j\in\mathcal N_i^{z}(t)} y_{ij}^{z}(t) \le 1,
\quad \forall i\in\mathcal M,\;\forall z\in\mathcal Z_i,\;\forall t.
\label{eq:association}
\end{equation}
Let $\gamma_i^{z\to z'}(t)\in\{0,1\}$ denote cross-technology  offloading of queued packets from interface $z$ to interface $z'$. 
At most one cross-technology migration is allowed per Alice per slot:
\begin{equation}
\sum_{z\in\mathcal Z_i}\sum_{\substack{z'\in\mathcal Z_i\\ z'\neq z}}
\gamma_i^{z\to z'}(t)\le 1,
\quad \forall i\in\mathcal M,\;\forall t.
\label{eq:offload_one_per_slot}
\end{equation}
Resource allocation decisions for bandwidth allocation $b_{ij}^{z}(t)\in[0,1]$ and power control $p_i^{z}(t)\in[0,1]$ satisfy
\begin{equation}
0 \le b_{ij}^{z}(t) \le y_{ij}^{z}(t),\qquad
\sum_{i\in\mathcal M}\sum_{j\in\mathcal N_i^{z}(t)} b_{ij}^{z}(t) \le 1,
\quad \forall z\in\mathcal Z,\;\forall t .
\label{eq:bw_constraints}
\end{equation}
\begin{equation}
0 \le p_i^{z}(t) \le 1,\qquad
P_{\mathrm{tx},i}^{z}(t)=p_i^{z}(t)\,P_{i,\max}^{z}\!\!\sum_{j\in\mathcal N_i^{z}(t)} y_{ij}^{z}(t).
\label{eq:power_constraints}
\end{equation}
where $P_{i,\max}^{z}$ is the maximum available power. 


The interface re-initialization decision $r_i^{z}(t)\in\{0,1\}$ initiates a $T^{\mathrm{rst}}$-slot re-initialization period when $r_i^{z}(t)=1$, during which interface $z$ remains silent:
\[
y_{ij}^{z}(\tau)=0,\qquad p_i^z(\tau)=0,\qquad
\forall j,\ \forall \tau\in[t,t+T^{\mathrm{rst}}-1].
\]
This models renewal of the interface identity or protocol context (e.g., MAC address randomization or security rekeying), 
which invalidates Willie’s interface-specific knowledge; its cost is captured implicitly through the resulting AoCI and delivery loss.

\subsection{Optimization Problem Formulation}

Let $\Delta_{i,z}(t)$ be the AoCI at the end of slot $t$ for interface ($i,z$).
$\mathcal{U}_{i,z}(t)$ denotes the set of packets originating from interface $z$ that are successfully delivered to any Bob during slot $t$.
If $\mathcal{U}_{i,z}(t)\neq\emptyset$, let $n^\star(t)$ be the freshest successfully delivered packet in that slot with generation time $t_{g,n^\star(t)}$. Then
\begin{equation}
\Delta_{i,z}(t)=
\begin{cases}
t - t_{g,n^\star(t)}, & \mathcal{U}_{i,z}(t)\neq\emptyset,\\[3pt]
\Delta_{i,z}(t{-}1)+1, & \mathcal{U}_{i,z}(t)=\emptyset,
\end{cases}
\label{eq:aoci_interface_update}
\end{equation}


with slot duration $\Delta t$.
The transmit energy on interface $z$ is
\begin{equation}
\label{eq:energy}
E_{i,z}(t) = P_{\mathrm{tx},i}^z(t)\,\Delta t,
\end{equation}
where $P_{\mathrm{tx},i}^z(t)$ is the transmit power.
\textcolor{black}{
Let $O_i^z(t)\in\{0,1\}$ denote the covert outage indicator, where 
$O_i^z(t)=1$ if at least one Willie monitoring protocol $z$ detects Alice $i$'s transmission in slot $t$. The aggregate covert throughput is
\begin{equation}
U(t) = \sum_{i\in\mathcal M}\sum_{z\in\mathcal Z_i(t)}
  y_{ij}^{z}(t)\, R_{ij}^{z}(t)\,
  \bigl(1-O_i^{z}(t)\bigr),
\label{eq:covert_throughput}
\end{equation}
where $R_{ij}^{z}(t)$ is the link capacity~\eqref{eq:rate}.
$U(t)$ equals 
the Shannon capacity on each active link when the covertness 
constraint~\eqref{eq:cov_power} is satisfied, and zero otherwise.
}
\textcolor{black}{
We use the reward
\begin{equation}
R(t) = w_u\sqrt{U(t)}
       - \!\!\sum_{i\in\mathcal M}\sum_{z\in\mathcal Z_i(t)}
         \!\!c_i^z(t),
\label{eq:reward_func}
\end{equation}
where 
$c_i^z(t) $ is the cost of interface $z$ in slot~$t$, and the concave utility $w_u\sqrt{U(t)}$ captures diminishing returns of throughput, capturing the trade-offs among covert reliability, exposure risk, information freshness, and energy consumption.
}

Define the packet covert delivery ratio and covert delivery efficiency as 
$\mathrm{PCR}^z(t)\triangleq
N_{\mathrm{cov},z}^{\mathrm{dlv}}/ N_{z}^{\mathrm{dlv}}$ 
and 
$\mathrm{CDE}^z(t)\triangleq
N_{\mathrm{cov},z}^{\mathrm{dlv}}/ N_{z}^{\mathrm{gen}}$, respectively.
The per-interface cost is
\begin{equation}
\begin{aligned}
c_i^z(t)
\triangleq\;&
  w_{\mathrm{pcr}}\bigl(\tau_{\mathrm{pcr}} - \mathrm{PCR}^z(t)\bigr)^{+} \\
& + w_{\mathrm{cde}}\bigl(\tau_{\mathrm{cde}} - \mathrm{CDE}^z(t)\bigr)^{+} \\
& + w_{\mathrm{exp}}\sqrt{N_{\mathrm{exp},i}^z(t)}
+ w_{\mathrm{aoi}}\,\Delta_{i,z}(t)
+ w_E\,E_{i,z}(t),
\end{aligned}
\label{eq:token_cost}
\end{equation}
where $(x)^{+} \triangleq \max(0,x)$. 
\textcolor{black}{
Successful DoS\_B and impersonation attacks reduce the number of successful covert transmissions (lowering PCR/CDE), increase retransmissions, and increase AoCI.
Bandwidth hijacking also reduces the available bandwidth for Alice's transmissions and increases AoCI.
Thus, minimizing (\ref{eq:token_cost}) encourages NC to improve covertness while promoting re-initialization decisions and data scheduling. 
}

\textcolor{black}{
We now formulate the long-term optimization problem that jointly optimizes the NC decisions
to maximize the reward~\eqref{eq:reward_func}. Specifically,
\begin{align}
(P1):\;
\max_{\mathbf r,\mathbf y,\boldsymbol{\gamma},\mathbf b,\mathbf p}
\;&
 \lim_{T\to\infty}\frac{1}{T}\,
\mathbb E\!\left[\sum_{t=0}^{T-1} R(t)\right]
\label{prob:P1}\\
\text{s.t.}\quad
& \eqref{eq:association}-\eqref{eq:energy}, \nonumber\\
& r_i^{z}(t), \, y_{ij}^{z}(t), \,  \gamma_i^{z\to z'}(t)\in\{0,1\}, \nonumber\\
& 0\le b_{ij}^{z}(t)\le 1,\; 0\le p_i^{z}(t)\le 1,\ \forall i,z,z',j,t. \nonumber
\end{align}
}

The problem (\ref{prob:P1}) is a stochastic optimization problem with coupled queueing and AoCI dynamics, 
and partial observability due to unobserved Willie strategies. Solving it exactly by re-optimizing the problem at every slot or via dynamic programming is computationally intractable. Therefore, in the next section, we reformulate it as a 
\textcolor{black}{POMDP}
and develop a PPO-based solution augmented with the proposed cross-domain multi-task lifelong learning framework.

\section{Cross-domain Multi-task Lifelong Learning}
\label{sec:lifelong}

The NC observes the network state but not Willie’s internal knowledge or attack feasibility; thus, we model the control problem as a POMDP
\(
\mathcal{P}=(\mathcal{S},\mathcal{A},\mathcal{O},\mathcal{T},R,\Omega).
\)
The state $s_t\in\mathcal S$ contains the NC-observable
system state $S_t^{\mathrm{NC}}$ (e.g., per-interface queue lengths and AoCI, link quality and interference), and hidden Willie internal variables that parameterize attack feasibility and behavior, including the knowledge variables, and the recency $T_{x,j}^{\mathrm{W,last}}$ of Willie’s knowledge about Bob $j$. This is the last-observation time when Willie $x$ successfully detected any Alice-Bob transmissions.

Given $s_t$ and the NC action, Willie attacks and their outcomes are generated stochastically through the transition function $\mathcal{T}$; the NC does not directly observe the realization of Willie actions and outcomes.
The NC action $a_t^{\mathrm{NC}}\in\mathcal A$ per slot $t$ is defined in Section~\ref{sec:datasched}.A, comprising Alice-Bob association, cross-technology offloading, bandwidth and power allocation, and interface re-initialization subject to their feasibility constraints in (\ref{eq:association})-(\ref{eq:power_constraints}).
The NC observation $o_t\in\mathcal O$ is given by its locally observable system state $S_t^{\mathrm{NC}}$, i.e., $o_t\triangleq S_t^{\mathrm{NC}}$. In particular, $o_t$ includes interface states and link/interference features available to the NC; Willie internal states are not observed. 
The transition $\mathcal{T}$ captures queue/AoCI evolution, mobility, channels, and Willie attack dynamics.
The reward is $\mathbb{E}[R(t)\mid s_t,a_t^{\mathrm{NC}}]$; and $\Omega$ is degenerate on $o_t$ because the NC deterministically observes $S_t^{\mathrm{NC}}$.

Observations are generated according to the conditional distribution
$o_t \sim \Omega(\cdot \mid s_t)$.
In our setting, the NC observes $S_t^{\mathrm{NC}}$ deterministically,
so $\Omega$ reduces to a degenerate distribution on
$o_t = S_t^{\mathrm{NC}}$, while Willie states remain unobserved.



\subsection{Learning Domains and Tasks}
\label{subsec:domain_task}

To capture the non-stationary environment faced by a multi-RAT UAV swarm while keeping learning tractable, we partition the experience stream into learning domains. Each domain aggregates scenarios with similar structural properties, which are expected to yield similar optimal control policies. 
We define each domain by a tuple $d=\big(d^{\mathrm{map}},d^{\mathrm{cov}},d^{\mathrm{port}} \big)$,
encoding map size, Willie coverage, and attack portfolio. 
The domains provide an interpretable abstraction for knowledge transfer by separating invariant structure from domain-specific variations, and enable effective parameter sharing across related tasks while permitting targeted adaptation when the environment changes. 
Task $n$ in domain $d$ is the POMDP $\mathcal{M}_{d,n}=(\mathcal{S},\mathcal{A},\mathcal{O},\mathcal{T}_{d,\xi_n},R,\Omega)$,
where $\xi_n$ denotes the adversary variables sampled at the start of the task (e.g., initial Willie knowledge and randomized initial conditions) and are fixed during the task. 
The NC observes only 
$o_t=S_t^{\mathrm{NC}}$, while Willie-side variables remain hidden 
and influence the dynamics through $\mathcal{T}_{d,\xi_n}$.
The resulting sequence of domain--task instances is denoted by $\{(d_n,\xi_n)\}_{n=1}^{N}$.

\subsection{PPO base learner for NC decisions}
\label{subsec:ppo_arch}

Proximal policy optimization (PPO)~\cite{schulman2017proximal} is used as the per-task optimizer for the NC POMDP. Each (Alice, protocol) pair is treated as a token. A shared
3-layer MLP encodes the observation tensor
$\mathbf{o}_t\!\in\!\mathbb{R}^{M\times Z\times F}$ into token
embeddings, and masked mean-pooling produces global and per-Alice
embeddings.
The actor uses five heads for reset, cross-technology offloading, Alice-Bob association, bandwidth, and power. 
Critic $V_\phi(o_t) $ \!=\! $v_\phi(\mathbf{e}^{\mathrm{glob}})$
is a shared-trunk MLP. 
Encoder and critic parameters $(\boldsymbol{\omega}, \boldsymbol{\phi})$ are shared across tasks; actor heads $\boldsymbol{\theta}_n$ are task-specific and factorized in
Section~\ref{subsec:factorized_param}.

\subsection{Cross-Domain Factorized Adaptive Lifelong Learning}
\label{subsec:factorized_param}

Queue, AoCI dynamics, and resource constraints are largely shared across domains, whereas the best policy trade-offs depend on topology, Willie monitoring profile, and attack regime. 
We therefore keep the encoder and critic shared and factorize only the actor heads. 
Following Cross-Domain PG-ELLA~\cite{ammar2015autonomous}, we write
\begin{equation}
\boldsymbol{\theta}_n = \boldsymbol{\Psi}^{(d_n)} \mathbf{L} \mathbf{s}_n,
\label{eq:theta_factorized}
\end{equation}
where $\mathbf{L}\in\mathbb R^{d_0\times K_{\mathrm{dict}}}$ is a shared knowledge base, $\boldsymbol{\Psi}^{(d)}\in\mathbb R^{p\times d_0}$ is a domain projection, and $\mathbf{s}_n\in\mathbb R^{K_{\mathrm{dict}}}$ is a task-specific coefficient. 
To reuse structure across related domains, we decompose $\boldsymbol{\Psi}^{(d)}$ into a shared base plus per-factor offsets:
\begin{equation}
\boldsymbol{\Psi}^{(d)} = \boldsymbol{\Psi}_0 + \Delta\boldsymbol{\Psi}_M[M(d)] + \Delta\boldsymbol{\Psi}_W[W(d)] + \Delta\boldsymbol{\Psi}_P[P(d)],
\label{eq:psi_componentwise}
\end{equation}
where $M(d)$, $W(d)$, and $P(d)$ index the map, Willie monitoring profile, and attack portfolio of domain~$d$, respectively. 
This additive parameterization is our structured domain projection and enables sharing across unseen factor combinations, while remaining task-level variation is absorbed by $\mathbf{s}_n$. 
Following LPG-FTW~\cite{mendez2020lifelong}, PPO updates $\mathbf{s}_n$ directly in the inner loop with proximal soft-thresholding, while $\mathbf{L}$ and $\boldsymbol{\Psi}^{(d)}$ are updated across tasks from stored summaries.
The lifelong optimization is
\begin{equation}
\begin{aligned}
\min_{\mathbf{L},\{\boldsymbol{\Psi}^{(d)}\},\{\mathbf{s}_n\}}
\quad
&\frac{1}{N}\sum_{n=1}^{N}\mathcal J_{d_n,\tau_n}(\boldsymbol{\omega}, \boldsymbol{\phi},\boldsymbol{\theta}_n)
+\mu\frac{1}{N}\sum_{n=1}^{N}\|\mathbf{s}_n\|_1 \\
&+\lambda_\mathbf{L}\|\mathbf{L}\|_F^2
+\lambda_{\boldsymbol{\Psi}}\sum_{d=1}^{D}\|\boldsymbol{\Psi}^{(d)}\|_F^2 \\
\end{aligned}
\label{eq:lifelong_objective}
\end{equation}

For outer updates, with anchor $\boldsymbol{\alpha}^{(n)}=\boldsymbol{\Psi}^{(d_n)} \mathbf{L} \mathbf{s}_n$, we use a quadratic surrogate:
\begin{equation}
\widehat \ell_n
=
\|\boldsymbol{\alpha}^{(n)}-\boldsymbol{\Psi}^{(d_n)} \mathbf{L} \mathbf{s}_n\|_{\mathbf{H}^{(n)}}^2
+{\mathbf{g}^{(n)}}^\top(\boldsymbol{\Psi}^{(d_n)} \mathbf{L} \mathbf{s}_n-\boldsymbol{\alpha}^{(n)}),
\label{eq:single_task_surrogate_cost}
\end{equation}
with $\mathbf{g}^{(n)}=\nabla_{\boldsymbol{\theta}} \mathcal J|_{\boldsymbol{\alpha}^{(n)}}$ and 
$\mathbf{H}^{(n)}\approx \tfrac{1}{2}\mathrm{diag}(\mathbf{F}_{\boldsymbol{\theta}}^{(n)})+\varepsilon_\mathbf{H} \mathbf{I}$.

The proposed algorithm to solve this problem, referred to as Cross-Domain Factorized Adaptive Lifelong Learning with PPO (CDFALL-PPO), is summarized in Algorithm 1.

\begin{algorithm}[t]
\scriptsize
\caption{\footnotesize{Cross-Domain Factorized Adaptive Lifelong Learning with PPO (CDFALL-PPO)}}
\label{alg:cross_domain_lpg_ftw}
\begin{algorithmic}[1]
\State Initialize shared $(\omega,\phi)$, dictionary $L$, domain projections $\{\boldsymbol{\Psi}^{(d)}\}$, memory $\mathcal M$.
\For{task instance $n=1,\dots,N$ with domain $d_n$}
\State Initialize sparse code $\mathbf{s}_n$; set $\boldsymbol{\theta}_n= \boldsymbol{\Psi}^{(d_n)} \mathbf{L} \mathbf{s}_n$.
\State \textbf{Inner loop:} update $(\boldsymbol{\omega}, \boldsymbol{\phi},\mathbf{s}_n)$ with PPO; apply proximal soft-threshold on $\mathbf{s}_n$.
\State Store task summary $(d_n,\mathbf{s}_n,\boldsymbol{\alpha}^{(n)},\mathbf{g}^{(n)},\mathbf{H}^{(n)})$ in $\mathcal M$.
\State \textbf{Outer loop:} sample mini-batches from $\mathcal M$ and update $\mathbf{L},\{\boldsymbol{\Psi}^{(d)}\}$ by gradient descent on \eqref{eq:single_task_surrogate_cost} plus Frobenius regularizers.
\EndFor
\end{algorithmic}
\end{algorithm}

\begin{table*}[t]
\centering
\scriptsize
\caption{Attack PCR and recovery PCR (RecPCR).} 
\label{tab:v1_implweighted_recpcr}
\begin{tabular}{p{4.1cm}p{1.5cm}p{1.4cm}p{1.0cm}p{1.0cm}p{3.0cm}p{2.8cm}}
\hline
Attack Task & 
CDFALL-PPO Attack PCR &
CDFALL-PPO RecPCR & 
PPO Attack PCR &
PPO RecPCR & 
Attack Dist under CDFALL-PPO (\%) (DoS/PH/BH/ACK) & 
Attack Dist under PPO (\%) (DoS/PH/BH/ACK) \\
\hline
DoS\_B + PH\_IB + BH\_IA + CPB\_ACK  & 0.70 & 0.93 & 0.52 & 0.64 & 18.6 / 12.7 / 9.5 / 59.3 & 24.0 / 17.9 / 14.0 / 44.0 \\
DoS\_B + PH\_IB + BH\_IA & 0.71 & 0.94 & 0.54 & 0.82 & 45.8 / 29.1 / 25.1 / 0.0  & 58.2 / 22.8 / 19.0 / 0.0 \\
DoS\_B only & 0.90 & 0.95 & 0.71 & 0.83 & 100.0 / 0.0 / 0.0 / 0.0  & 100.0 / 0.0 / 0.0 / 0.0 \\
\hline
\end{tabular}
\end{table*}




\section{Simulation Results}

\begin{figure}
    \centering
    \includegraphics[width=\linewidth]{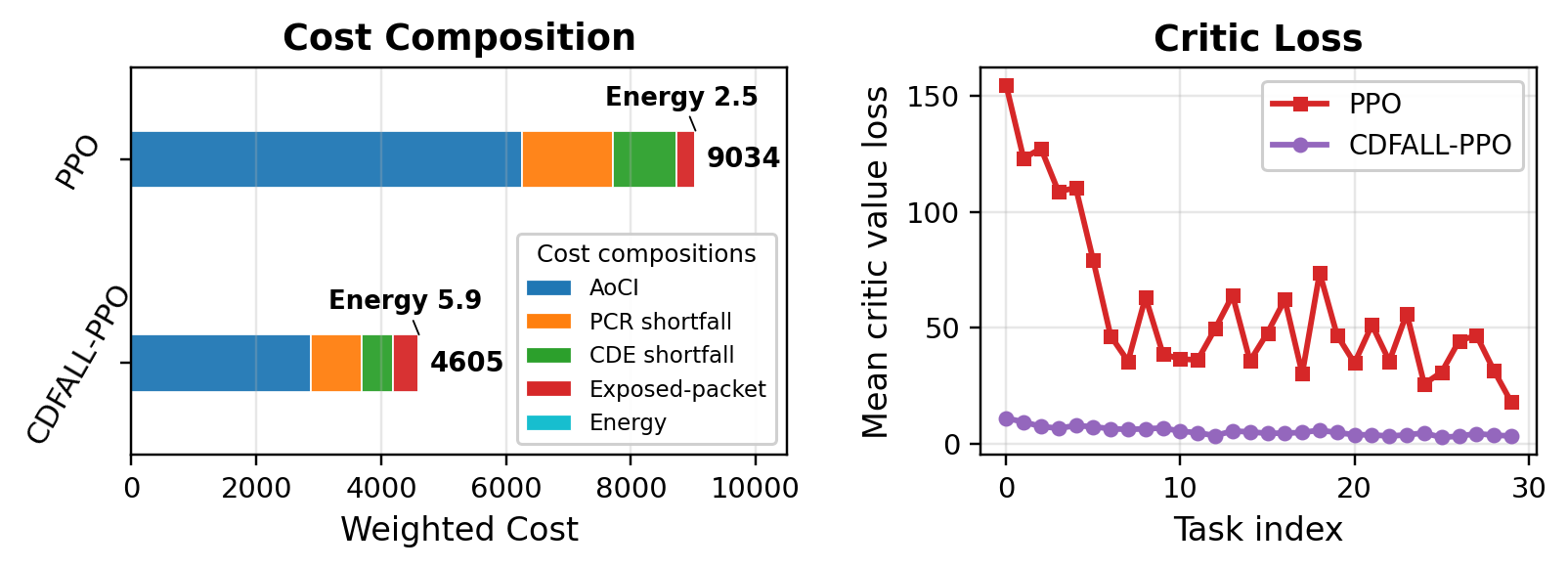}
    \caption{Cost decomposition and mean critic value loss per task.}
    \label{fig:reward_critic}
\end{figure}


\begin{figure}
    \centering
    \includegraphics[width=1\linewidth]{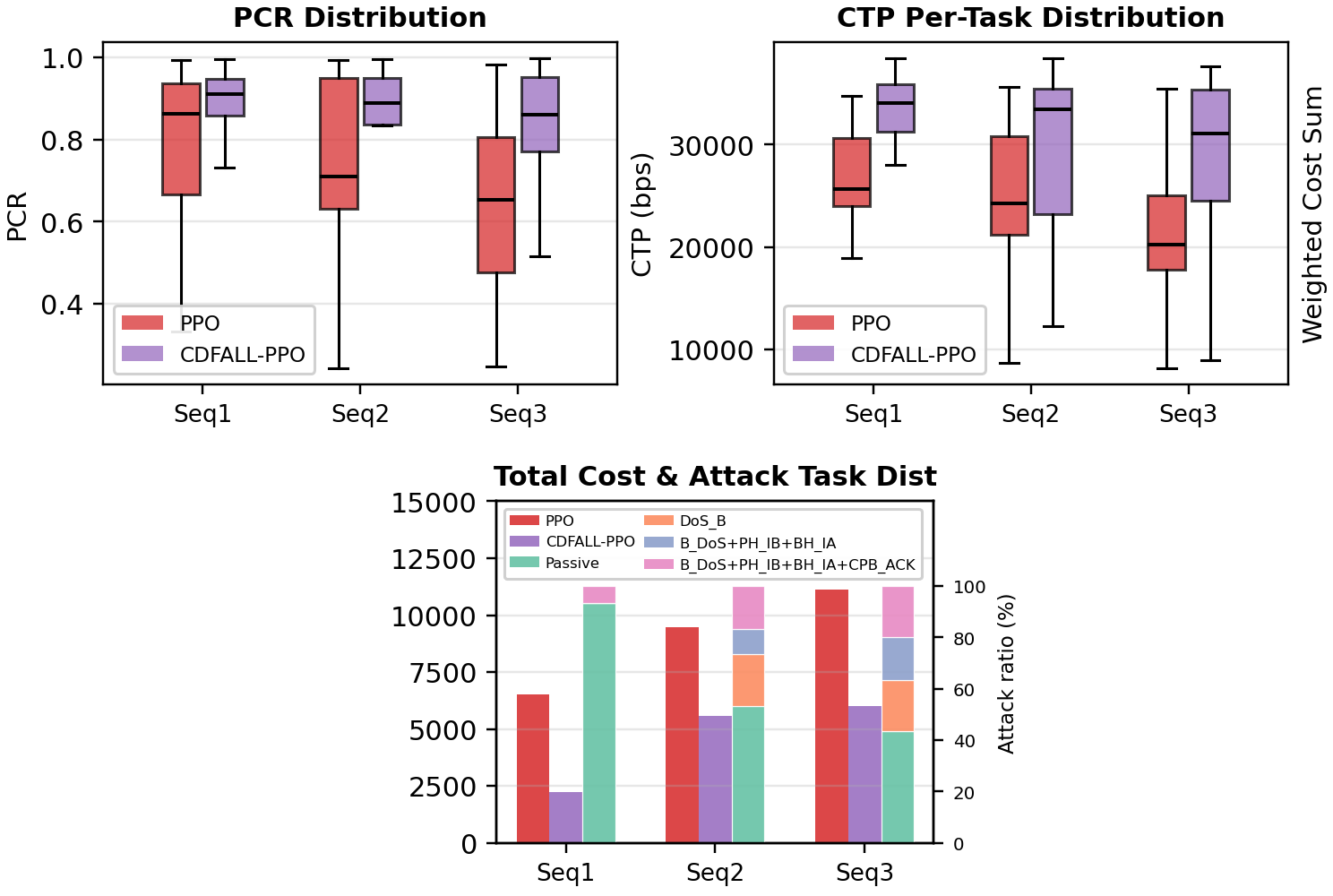}
    \caption{Per-task PCR and CTP distributions, weighted cost, and Willie attack-type composition across three 30-task sequences. }
    \label{fig:pcr_ctp_cost_seq1to3}
\end{figure}

\subsection{Experimental Setup}
\label{subsec:exp_setup}






We conducted extensive simulations in Python to assess the performance of our proposed scheme. We evaluate CDFALL-PPO against PPO on three escalating 30-task sequences (Seq1--Seq3), generated randomly under increasing threat regimes.
%
%
Each domain is indexed by topology $d^{\mathrm{map}}$, represented by map size $L^{\mathrm{map}}\!\times\!L^{\mathrm{map}}$ area with $L^{\mathrm{map}}\in\{800,1200,1600\}$, 
Willie monitoring profile $d^{\mathrm{cov}}$ indicating the subset of RATs monitored (e.g., \texttt{wifi\_only}, \texttt{wifi\_5g}, \texttt{all}), and attack portfolio $d^{\mathrm{port}}$, ranging from passive detection to the composite set DoS\_B+BH\_IA+CPB\_ACK+PH\_IB.

We design three 30-task sequences with progressively escalating
threats.
\textit{Seq\,1} uses a single 800\,m x 800\,m map with 26 of 30 tasks
under passive-only Willie monitoring (portfolio\,0); the remaining 4
tasks introduce single-protocol active attacks (portfolios\,1--2),
providing a low-conflict baseline.
\textit{Seq\,2} balances all four threat portfolios
(0--3) across the three map sizes and four Willie
coverage types, creating high inter-task domain diversity.
\textit{Seq\,3} deliberately interleaves anchor tasks
with abrupt \emph{shock} transitions---all portfolios appear,
including full CPB\_ACK escalation (portfolio\,3)---using six coverage
types and anchor-shock-revisit cycles that maximise catastrophic
interference.
Every task specifies a covertness parameter $\epsilon{=}0.1$
($\rho_\varepsilon{\approx}0.35$),
an SINR-based Willie decode threshold of $-10$\,dB,
and 1--2 Bobs per protocol. 

We assume
$\bar p_{\mathrm{BH\_IA},\text{WiFi}}^{\mathrm{fea}}{=}0.5$ relative to the 75\% MAC de-randomization reported in~\cite{matte2016timing},
while $\bar p_{\mathrm{BH\_IA},\text{5G}}^{\mathrm{fea}}{=}0.05$ reflects its mutual authentication.
For low-power IoT protocols, Choi et al.~\cite{choi2016dissecting}
report 48.19\% automated spoofing success on IEEE~802.15.4,
calibrating our ZigBee ($0.30$) and LoRa ($0.35$) values
as conservative lower bounds.
Each task uses three UAV Alices at an altitude $100$\,m, each equipped with
WiFi, 5G NR, LoRa, and ZigBee interfaces.
Willie density is
$\lambda_w\in\{0.01,0.02,0.03\}$, yielding 1--5 Willies depending on map
size and monitored protocols. The slot duration is $0.5$\,s, episodes
last 200 slots, and interface reset lasts $5$, $20$, or $50$ slots. PPO
uses learning rate $3\times10^{-4}$, $\gamma=0.99$, GAE $\lambda=0.95$,
clip $0.2$, and 4 epochs/update. CDFALL-PPO uses $d_0=96$,
$K_{\mathrm{dict}}=48$, task-code learning rate $\eta_s=0.03$, and
sparsity weight $\mu=10^{-3}$. Both methods use the same reward weights
$w_{\mathrm{PCR}}=2.4$, $w_{\mathrm{CDE}}=2.0$,
$w_{\mathrm{Exposed}}=0.9$, $w_{\mathrm{AoCI}}=0.18$,
$w_{\mathrm{Energy}}=0.2$, and $w_{\mathrm{Util}}=0.08$.

\subsection{Numerical Results}
\label{subsec:cdll_results}

We evaluate CDFALL against PPO on the three 30-task sequences described above.
Each algorithm--sequence pair is trained with three independent seeds (100, 101, 102) for $512$ environment steps per task.

\paragraph{Cost decomposition and optimization stability}
Fig.~\ref{fig:reward_critic} decomposes weighted cost terms on Seq1, averaged over Seq1--3. CDFALL-PPO achieves half the total weighted cost of PPO (4605 vs.\ 9034), with the largest savings in AoCI and PCR. The critic loss of PPO oscillates between 30--150 throughout training, whereas CDFALL-PPO converges below 15 after the first few tasks.
Compared with PPO, CDFALL-PPO reduces AoCI cost by 72.0\%, PCR penalty by 67.3\%, and CDE by 68.6\%.
The critic value-loss trajectory is also lower overall (task-wise mean: 1.08 vs. 3.19), consistent with more stable optimization under non-stationary tasks. 
\textcolor{black}{
These gains stem from reusable knowledge shared via $\mathbf{L}$, while the factorized projection $\boldsymbol{\Psi}^{(d)}$ recombines map-, coverage-, and attack-specific components to enable positive transfer. Relative to PPO, this reduces interference and catastrophic forgetting, ultimately lowering critic oscillation and costs.
}

\textcolor{black}{
\paragraph{Attack robustness and post-attack recovery}
Table~\ref{tab:v1_implweighted_recpcr} shows that CDFALL-PPO is stronger
both on the tasks with active attacks and on the immediately following recovery tasks.
Under the full composite attack set
(DoS\_B+PH\_IB+BH\_IA+CPB\_ACK), Attack PCR rises from 0.52 to 0.70 and
RecPCR from 0.64 to 0.93. Under DoS\_B+PH\_IB+BH\_IA, the gains remain
substantial (0.54 to 0.71 on tasks with active attacks and 0.82 to 0.94 on the next
task). Even for DoS-only tasks, CDFALL-PPO improves on-attack PCR from 0.71
to 0.90 and post-attack PCR from 0.83 to 0.95. The largest recovery gap appears
when CPB\_ACK is present: under CDFALL-PPO, Willie devotes a larger share
of its attack mix to CPB\_ACK and a smaller share to direct DoS/PH/BH
than under PPO, consistent with more Willie effort spent on
reconnaissance before effective exploitation. Overall, CDFALL-PPO both
reduces damage during attack tasks and restores high PCR more reliably on
the next task.
}

\paragraph{Performance under escalating cross-domain threats}
Fig.~\ref{fig:pcr_ctp_cost_seq1to3} summarizes PCR, CTP distributions, 
aggregate weighted cost, and Willie attack-task composition across Seq1--Seq3.  The fraction of active-attack tasks rises from 7\% (Seq1) to 47\% (Seq2) and 57\% (Seq3), with
increasingly complex multi-vector attacks (DoS, \textcolor{black}{impersonation of Alice/Bob}, CPB-ACK).
CDFALL-PPO's median PCR advantage over PPO widens from
+5\,pp (91\% vs.\ 86\%) in Seq1 to +18\,pp (89\% vs.\ 71\%) in Seq2
and +21\,pp (86\% vs.\ 65\%) in Seq3, and median CTP gains grow from
+8\,396 to +9\,146 to +10\,842\,bps accordingly.
Meanwhile, the aggregate weighted cost is about half that of PPO in
every sequence.
The results indicate that the
performance gap is \emph{threat-driven}: as attack pressure intensifies,
PPO's PCR collapses (median drops from 86\% to 65\%) while CDFALL-PPO
degrades gracefully (91\% to 86\%), retaining its learned covert
policies across domain shifts. 
\textcolor{black}{The factorized basis $\boldsymbol{\Psi}^{(d)}$ isolates threat-regime variations to domain-specific adapter offsets, so knowledge learned under different attacks is preserved via the shared library $\mathbf{L}$ rather than being overwritten by subsequent tasks. Consequently, CDFALL-PPO reduces the catastrophic forgetting and prevents PCR collapse observed with PPO as attacks escalate.
}

\section{Conclusion}
This paper studied covert data offloading and covertness recovery in multi-RAT UAV swarms under post-detection Willie threat escalation. 
We jointly optimized Alice-Bob association, RAT selection, bandwidth and power allocation, and interface re-initialization under adaptive wardens. We model wardens' attack-selection process, capturing escalation and nonstationary protocol-aware strategies. To restore covertness across different environments with varying Willie densities and attack types, we proposed CDFALL-PPO, a cross-domain factorized lifelong RL framework that adapts online while mitigating catastrophic forgetting. Simulations show CDFALL-PPO significantly improves PCR (on-attack and post-attack), AoCI, and cost compared to existing baselines.




\bibliographystyle{IEEEtran}
\bibliography{references}

\end{document}